\documentstyle[preprint,aps,epsfig]{revtex}

\def\bme{\mbox{\boldmath$\varepsilon$}}

\begin{document}

\draft
\preprint{
\vbox{
\hbox{ADP-98-25/T301}
}}

\title{Light-Cone Meson-Baryon Fluctuations and Single-Spin 
Asymmetries} 

\normalsize
\author{C. Boros} 
\address {Department of Physics and Mathematical Physics,
                and Special Research Center for the
                Subatomic Structure of Matter,
                University of Adelaide, 
                Adelaide 5005, Australia}

\date{\today}
\maketitle

\begin{abstract} 

We show that energetically favored  meson-baryon 
fluctuations present in the light-cone wave function 
of a transversely polarized proton 
can account for the 
left-right asymmetries measured in inclusive meson production 
processes.  
\end{abstract}  

\tighten
\newpage

Recently, remarkable left-right asymmetries have been measured 
in inclusive meson production using transversely
polarized projectile hadrons and unpolarized hadron targets \cite{FNAL}.  
It is observed that these  asymmetries have 
many striking features characteristic of leading particle 
production. They are significant in and only in the 
fragmentation region of the transversely polarized projectile, 
they depend on the quantum numbers both of the projectile and 
of the produced particles, but are insensitive to the 
quantum numbers of the target.

While considerable amount of experimental information 
has  been collected over the last few years \cite{FNAL}, there is 
still no accepted theoretical concept for the explanation 
of the experimental findings \cite{Boros98}.  
Since such asymmetries are expected to vanish 
because of  helicity conservation of the almost massless 
quarks \cite{Kane} in leading twist, leading order 
QCD it has become widely accepted that 
modeling of higher twist, nonperturbative  effects is needed 
to understand these phenomena. 
Here, we make such an attempt and show that 
energetically favored meson-baryon fluctuations 
present in the light-cone wave function 
of the projectile, can account for the nonvanishing 
left-right asymmetries in the projectile fragmentation 
region.

It is instructive to review briefly  the connection between 
light-cone fluctuations  and  leading particle production.  
Mesons produced in the fragmentation regions in hadron-hadron 
collisions are known to  
reflect the quantum numbers of the projectile and,  
thus, carry information about its wave function. 
It has been realized that the production of leading 
mesons 
for moderately large transverse momenta, 
especially the flavor asymmetry between leading and nonleading 
particle production, can be understood on the basis of 
higher Fock states present in the light-cone wave functions 
of hadrons\cite{Vogt}. In this picture, the projectile can fluctuate 
into higher Fock states containing quark-antiquark pairs 
in addition the valence quarks which carry its quantum 
numbers. 
The most probable fluctuations  
are those which have  minimal invariant mass.  
Coalescence of the antiquark  with 
valence quarks of the projectile produces then leading 
particles in the projectile fragmentation region
\cite{Vogt,Hwa}. Alternatively, 
one can think of leading particles as preexisting  
in the higher Fock states  of the projectile 
in form of energetically favored meson-baryon fluctuations 
\cite{Vogt,Thomas,Speth}.  
Then, production of leading particles happens when 
soft interactions between 
the projectile and target break the coherence of the 
light-cone meson-baryon fluctuation and bring the particles 
on shell. 
Since these soft interactions  involve small momentum transfers between 
target and projectile  the momentum distribution of the produced leading 
particles should closely resemble the momentum distribution 
of the Fock state (light-cone wave function) \cite{Vogt,Hwa}.  
Since the transverse momenta of the produced particles 
in the single-spin experiments are typically $p_\perp\sim 0.7$ GeV/c,  
a picture based rather on meson-baryon than on quark 
degrees on freedom should be more appropriate for the description 
of the asymmetries.

Let us recall the definition of the left-right asymmetries 
for the reaction $p(\uparrow ) + p(0)\rightarrow M + X$,  
where $p(\uparrow )$ and $p(0)$ stand for the transversely 
polarized projectile and unpolarized target protons, respectively;     
$M$  stands for the inclusive produced meson and $X$ for the unobserved 
final states:  
\begin{equation} 
 A_N(x_F|M ) =\frac{N_L(x_F ,\uparrow |M) - 
N_L(x_F  ,\downarrow |M)}{N_L(x_F ,\uparrow |M) + 
N_L(x_F ,\downarrow |M)},  
\label{an}  
\end{equation} 
where
\begin{equation}
   N_L(x_F,i|M)=\frac{1}{\sigma_{in}} 
\int_{(D)} d^2p_\perp \frac{d\sigma (x_F,{\bf p}_\perp,i|M)}
{d x_F d^2p_\perp}  
\label{def}
\end{equation} 
($i=\uparrow$ or $\downarrow$) is the normalized number 
density of the produced mesons observed in a given 
kinematical region $D$ ($p_\perp >0.7$ GeV/c for example) 
on the left-hand side of the beam ($L$) looking down stream. 
$\sigma_{in}$ is the total inelastic cross section, 
$x_F$ is the usual Feynman-$x$  $x_F=2p_{\parallel}/\sqrt{s}$, where  
$p_\parallel$, $p_\perp$ are the longitudinal and transverse 
momenta of the produced mesons and $\sqrt{s}$ is the c.m. energy.  
Since $N_L(x_F,\uparrow |M)=N_R(x_F,\downarrow |M)$ and 
$N_L(x_F,\downarrow |M)=N_R(x_F,\uparrow |M)$, where $N_R(x_F,i|M)$ 
are the corresponding number densities measured on the 
right-hand side, it is 
clear why $A_N$ is usually referred to as left-right asymmetry. 

According to this definition,  
nonvanishing left-right asymmetries reflect a remarkable 
correlation between the direction of transverse motion  
of the produced particles and the transverse polarization of the 
projectile. In  the above picture for leading particle 
production this correlation is expected to be {\it  present} in 
the wave function of the projectile and  to be carried over to the 
leading particle appearing in the final state. 
Thus, left-right asymmetries resemble rather a correlation 
{\it preexisting} in the initial state than a correlation produced 
in the final state, for example,  by fragmentation.  
We note that this conjecture is in agreement with recent 
experimental observations by the Tasso \cite{Tasso}, Aleph \cite{Aleph} and 
SLD Collaborations \cite{SLD} showing no transverse polarization 
of the produced particles in $e^+e^-$ annihilations.

In pion production the lowest Fock 
states involving $u\bar u$ and $d\bar d$ 
fluctuations are $p(uud)=\pi^+(u\bar d) n(udd)$, 
$p(uud)=\pi^0[\frac{1}{\sqrt{2}}(u\bar u + d\bar d)] p(uud)$ and   
$p(uud)=\pi^-(d\bar u) \Delta^{++}(uuu)$, respectively.  
Since  higher mass fluctuations have relatively small 
probabilities, the energetically favored lowest fluctuations play 
the most important role. However, 
for the sake of completeness, we also include the relatively 
less important $\pi^+$ and $\pi^-$ fluctuations 
$p(uud)=\pi^+(u\bar d)\Delta^0 (udd)$ and 
$p(uud)=\pi^0[\frac{1}{\sqrt{2}}(u\bar u + d\bar d)] \Delta^+(uud)$.     
While the lowest lying meson-baryon fluctuations for $\pi^+$ and $\pi^0$ 
contain  spin-1/2 baryons the lowest  fluctuation for 
$\pi^-$ contains the spin-3/2 baryon $\Delta^{++}$.   
This will be crucial for understanding the left-right asymmetries. 
Therefore, let us examine the  wave functions of 
the meson-baryon fluctuations containing spin-1/2 and spin-3/2 
baryons, respectively. 
The form of the wave function follows from the requirement 
that the total quantum numbers of the proton must be 
conserved.  Since  pions and kaons are pseudoscalar particles with negative 
parity and the baryons involved have the same parity 
as the proton the meson-baryon system must have odd 
angular momentum. Thus, the total angular momentum part of 
the wave functions, $\Psi^{MB(S)}_{J,J_z,L,S}$  
in the center of mass reference frame of the pseudoscalar 
meson-baryon ($MB(S)$) fluctuation containing a spin $1/2$-baryon 
$B(\frac{1}{2})$ is given by 
\begin{equation} 
  \Psi_{\frac{1}{2},
 \frac{1}{2}1,\frac{1}{2}}^{MB(\frac{1}{2})} 
   =  \sqrt{\frac{2}{3}} \psi_{1}^{+1} \, 
                         \chi_\frac{1}{2}^{-\frac{1}{2}}
     -\sqrt{\frac{1}{3}} \psi_{1}^{\,0} \, 
                         \chi_\frac{1}{2}^{+\frac{1}{2}}
\label{eq3} 
\end{equation}
Here, $J$, $L$, $S$ and $J_z$, $L_z$, $S_z$ are the total 
angular momentum, total orbital angular momentum and total spin 
of the meson-baryon fluctuation 
and their $z$ components, respectively.   
$\psi_{L}^{L_z}$ and $\chi_{S}^{S_z}$ are the 
total orbital-angular momentum and total spin part of the 
wave function.  
It follows that  the $M B(\frac{1}{2})$ fluctuation is predominantly 
in a state with positive orbital angular 
momentum $L_z=1$ in a transversely upwards polarized proton. 
Thus, the meson and the baryon  perform orbital 
motion around their center-of-mass; since the baryons 
involved are much heavier than   
the mesons (pions and kaons),  
it is essentially the meson which ``orbits'' around 
the baryon. Hence, one can speak of effective  ``pion currents'' 
in such Fock states. This current is  is anticlockwise  
with respect to the polarization axis ($z$ axis) 
of the transversely polarized proton  for 
pseudoscalar mesons ($M$) in a $M B(\frac{1}{2})$ state,  
since the $L_z=1$ component dominates. 

On the other hand, the total angular momentum part of the 
wave function of the meson-baryon fluctuation   
of an upwards polarized proton containing a spin-3/2 baryon 
$B(\frac{3}{2})$ and a pseudoscalar mesons $M$  is given by 
\begin{equation} 
  \Psi_{\frac{1}{2},\frac{1}{2},1,\frac{3}{2}}^{MB(\frac{3}{2})} 
   =  \sqrt{\frac{1}{2}} \psi_{1}^{-1}\, 
                         \chi_\frac{3}{2}^{+\frac{3}{2}}
     -\sqrt{\frac{1}{3}} \psi_{1}^{\,0}\,  
                         \chi_\frac{3}{2}^{+\frac{1}{2}}
     +\sqrt{\frac{1}{6}} \psi_{1}^{+1}\,  
                         \chi_\frac{3}{2}^{-\frac{1}{2}}
\label{eq4} 
\end{equation}
We see that the $MB(\frac{3}{2})$ system is more likely to have negative 
than positive  orbital angular momentum. Thus, pseudoscalar 
mesons in these Fock states perform  orbital motion mainly  
clockwise with respect to the polarization axis
of the transversely polarized proton.  

In order to understand the significance of these observations, we 
discuss  the formation of the leading mesons. It is convenient 
to work in the center of mass  of two particle system. 
In this system, the projectile is an extended object and 
the target is  Lorentz contracted. 
Soft interaction between the projectile and target  
become effective when the target and the projectile partially  
overlap. In leading meson production,   
the meson is expected to be  brought on mass 
shell through soft interaction with the target.  
If this happens on the front surface of the  
extended baryon-meson state, i.e., when the target begins to overlap 
with the target,  the freed meson will have a larger 
probability to ``go'' left or to ``go'' right 
if it belongs to a $MB(\frac{1}{2})$ or $MB(\frac{3}{2})$ fluctuation. 
This is  because, on the front surface,  $L_z=1$ ($L_z=-1$) of the 
$MB(\frac{1}{2})$ ($MB(\frac{3}{2})$) system 
means  that  the probability for the pion to have  
transverse momenta pointing to the left (to the right) is larger than  
to have one pointing to the right (to the left)  \cite{Meng}. 
However, when this happens on the back surface, i.e. after 
the target  went through the projectile, 
the soft interactions  of the spectator quarks with the target 
destroy  the coherence of the Fock state and the 
produced mesons cannot retain their preferred transverse direction.  
Our expectation that initial state interactions  
with  spectator quarks in leading particle production  
plays an important role agrees with the observation that  
leading particle  
production occurs dominantly when the spectator quarks  
interact strongly in the target, leading to  strong 
nuclear dependence \cite{Vogt}.   
Preliminary results from the E704 Collaboration showing a strong 
$A$ dependence of $A_N$ are also consistent with such an 
expectation \cite{Plat}

Since the leading $\pi^+$ and $\pi^0$ are mainly 
produced by light-cone fluctuations 
containing spin-1/2 baryons, we expect positive 
left-right asymmetries  for both  $\pi^+$ and $\pi^0$. 
On the other hand, the asymmetry for $\pi^-$ should be negative, since,  
here, the lowest lying Fock state contains a spin-3/2 baryon. 
Furthermore, the remarkable ``mirror'' symmetry  
of the left-right asymmetries between  transversely polarized 
proton and antiproton projectiles follows immediately 
from the proposed picture. This is because 
the lowest lying Fock states of the antiproton 
containing $\pi^+$ and $\pi^-$ are 
$\bar p(\bar u\bar u\bar d) = \pi^+(u\bar d) 
\bar\Delta^{++}(\bar u\bar u\bar u)$ and  
$\bar p(\bar u\bar u\bar d) = \pi^-(d\bar u) 
\bar n(\bar u\bar d\bar d)$, respectively. Thus, in contrary to 
the situation in $pp$ collision, 
the partner of the $\pi^+$ 
is  a spin-3/2  and that of the  $\pi^-$ is a spin-1/2 baryon. 
Therefore, we expect positive 
left-right asymmetry for $\pi^-$ and negative asymmetry for 
$\pi^+$. Note that the asymmetry for $\pi^0$ remains 
unchanged. Furthermore, we also expect to see  left-right 
asymmetries for $\eta$ meson production. These asymmetries 
should be  similar in magnitude 
to that in $\pi^0$ productions and positive for both 
proton and antiproton projectiles.  
All these features have been observed  
experimentally \cite{FNAL}.     
  
Encouraged by the good qualitative agreement with the 
experimental data we want to describe the asymmetries 
quantitatively.  
While leading  mesons are predominantly produced 
by  meson-baryon fluctuations and populate the large 
$x_F$ region,  nonleading mesons 
are  produced by other mechanisms such as string 
fragmentation, higher Fock states and populate the central 
rapidity region. For large enough transverse momenta, the production 
of mesons in the central region (and also in the 
fragmentation regions)  can be described by leading twist perturbative QCD. 
However, since   the typical  transverse momenta involved in 
left-right asymmetry measurements are in the 
order of $p_\perp\sim 0.7$ GeV/c, we can not expect 
pQCD to be applicable and factorization 
to be valid.  Therefore, in the following we 
calculate the spectra in the projectile fragmentation region using 
the light-cone wave functions of the meson-baryon fluctuation and  
use a fit to the 
nonleading spectra of the produced mesons. 
In  separating the leading from the nonleading 
spectrum, we can use kaon-production as guideline. 
This is because $K^-$ has no common valence quarks with the projectile 
proton, i.e. there are no leading $K^-$ in proton-proton collisions. 
Thus, the inclusive $K^-$ spectrum is  
equivalent to the nonleading spectrum of $K^+$.   
Assuming that the {\it form} of  the nonleading spectrum is  
independent of the produced mesons, we can use  
exact the same form as obtained by fitting the $K^-$ spectrum 
for the nonleading parts of  the pion production  
allowing for a normalization \cite{Boros96}.

In the meson-baryon fluctuation model, the differential cross section 
for producing leading mesons is given by 
\begin{equation} 
  \frac{d\sigma_{MB}}{dx_Fd^2 p_\perp}  
 = \sigma_{pB}\frac{1}{2} \sum_{Ss} 
\frac{d P^{Ss}_{MB}}{dy d^2 p_\perp} 
\delta(x_F-y).    
\label{cross} 
\end{equation} 
Here, $\sigma_{pB}$ is the total inclusive 
proton-baryon inelastic cross section at c.m. energy of 
$s(1-x_F)$. We use $\sigma_{pp}$ and the numerical 
value $40$ $mbarn$ for $\sigma_{pB}$.     
$dP^{Ss}_{MB}/dyd^2 p_\perp$ is the probability that the proton  
fluctuates in a meson and baryon   
with longitudinal momentum fractions  $y$ 
and $1-y$, transverse momentum ${\bf p}_\perp$ and 
$-{\bf p}_\perp$, respectively.    
$S$ and $s$ denote the spin projections of the proton and the 
baryon with respect to a conveniently chosen axis.  
In the following we use both a nonrelativistic toy model and   
the meson-cloud model to calculate these probabilities.    

In a nonrelativistic approximation, 
the probabilities  for different angular momenta 
are simply related by Clebsch-Gordan coefficients as 
given by Eq. (\ref{eq3}) and 
in Eq. (\ref{eq4}). For the  spin-independent momentum-space part, we use  
a simple phenomenological ansatz, used in Ref. 
\cite{Brodsky96},   
\begin{equation} 
\frac{dP_{MB}}{dyd^2p_\perp}({\cal M}) = 
A_{MB} (1+{\cal M}^2/\alpha^2)^{-p} . 
\label{wave} 
\end{equation}
Here, ${\cal M}$ is the invariant mass of the meson-baryon 
fluctuation 
${\cal  M}^2 =(p_\perp^2 + m_M^2)/y + (p_\perp^2 + m_B^2)/(1-y)$ and  
$m_M$ and $m_B$ are the meson and baryon masses.   
$A_{MB}$ is a normalization constant, for the other parameters,  
we use the values, $\alpha =330$ MeV and $p=3.5$ Ref.\cite{Brodsky96}. 
The relative weights of  fluctuations containing  proton, neutron and 
Delta baryons are given by the  isospin factors: 
$(\pi^+ n):(\pi^0 p):(\pi^+ \Delta^0) :( \pi^0 \Delta^+):  
(\pi^-\Delta^{++})=2:1:\frac{1}{3}:\frac{2}{3}:1$ \cite{Kumano}.  
We fixed the  normalization  by  
fitting the $\pi^-$ cross section. Then,  
the probabilities for the above fluctuations, 
are  $16\%$, $8\%$, $0.8\%$, $1.6\%$, and $2.4\%$, respectively.            
The  probability for the $K^+\Lambda$ fluctuation is $1.7\%$.          
These probabilities have approximately the right magnitude 
to account for the Gottfried sum rule violation 
\cite{Thomas2,Henley,Kumano}  
measured by the New Muon Collaboration (NMC) \cite{NMC}.   
We calculated the invariant cross sections   
$Ed\sigma/d^3p=\frac{x_F}{\pi}d\sigma/dx_Fdp_\perp^2$ 
for pion and kaon production  
as a function of $x_F$ at $p_\perp =0.75$ GeV/c. 
The result is shown in Fig.1a-c together with  the data 
\cite{CHLM}.

In the meson-cloud model, the probabilities in Eq. (\ref{cross}) 
are given by    
\begin{equation} 
   \frac{dP^{Ss}_{MB}}{dyd^2 p_\perp} 
 = \frac{g^2_{NBM}}{16\pi^3}  
    \frac{|{\cal T}_{MB}^{Ss}|^2}{y(1-y)}  
    \frac{|G_{NMB}(y,p_\perp^2)|^2}{
 (m_N-{\cal M}^2_{MB})^2} \,\,. 
\end{equation} 
Here, $|{\cal T}_{MB}^{Ss}|^2$ 
contains  the spin-dependence of the 
probabilities and can be obtained by calculating traces over nucleon and 
baryon spinors \cite{Thomas,Speth}.   
$G_{NMB}(y,p^2_\perp )$ are phenomelogical vertex functions. 
They are often parametrized in a power form 
\begin{equation} 
     G_{NMB}(y,p_\perp^2 )= 
\left(\frac{\Lambda_{MB}^2+m_N^2}{\Lambda_{MB}^2+{\cal M}^2_{MB}}\right)^n.  
\end{equation} 
We found a good fit to the data by choosing $n=3$  and  
$\Lambda_{\pi N}, \Lambda_{\pi\Delta},\Lambda_{K\Lambda}=1.6,1.25,1.7$ 
GeV.    
The coupling constants are $g^2_{pp\pi^0}/4\pi =13.6$, 
$g^2_{p\Delta^{++}\pi^-}/4\pi =12.3$ GeV$^{-2}$ and 
$g^2_{p\Lambda K^+}/4\pi =14.7$. 
The results  are  shown in Fig. 2.  

In order to calculate the left-right asymmetries we note that 
only the leading particles will contribute to the asymmetry.  
In the nonrelativistic toy-model,     
the probability, 
$P_{MB}^{J_z,L_z}(x_F,{\bf p}_\perp )
\equiv dP_{MB}^{J_z,L_z}/dx_F d^2p_\perp$, 
for a meson-baryon fluctuation containing a  
baryon with spin $s$  to be in an orbital angular 
momentum state with $L_z$, when the projectile has total angular momentum 
projection, $J_z$,  is equal to the unpolarized 
probabilities, $P_{MB}(x_F,p_\perp )$ multiplied by appropriate 
Clebsch-Gordan coefficients. 

According to the proposed picture, only mesons produced 
by $MB$ fluctuation contribute to the 
difference of the measured number densities,  
$\Delta N_L(x_F,{\bf p}_\perp |M)\equiv 
N_L(x_F,{\bf p}_\perp , \uparrow |M) 
-N_L(x_F,{\bf p}_\perp , \downarrow |M)$. 
Obviously, this quantity should be proportional 
to $\Delta P_{MB}(x_F,{\bf p}_\perp ) \equiv      
P^{\frac{1}{2},1}_{MB}(x_F,{\bf p}_\perp )-
P^{\frac{1}{2},-1}_{MB}(x_F,{\bf p}_\perp )$,  
the difference of the probabilities for the $MB$ fluctuations  
to have $L_z=1$ and $L_z=-1$ in an upwards polarized proton. 
(Note that we have the relations 
$P^{J_z,L_z}_{MB}(x_F,{\bf p}_\perp )=
P^{-J_z,-L_z}_{MB}(x_F,{\bf p}_\perp )$  and 
$P^{J_z,-L_z}_{MB}(x_F,{\bf p}_\perp )=
P^{-J_z,L_z}_{MB}(x_F,{\bf p}_\perp )$).    
The proportionality constant,  $C$ ($0\le C\le 1$),      
between $\Delta N_L(x_F,{\bf p}_\perp |M)$ and 
$\Delta P_{MB}(x_F,{\bf p}_\perp )$, 
effectively describes how strong  
spectator quarks interact in the target. 
The sum of the number densities 
$N_L(x_F,{\bf p}_\perp )=N_L(x_F,{\bf p}_\perp , \uparrow |M) 
+N_L(x_F,{\bf p}_\perp , \downarrow |M)$ is 
twice the corresponding unpolarized number density 
measured on the left-hand side and is given by  
$N^{nl}(x_F,p_\perp )+P_{MB}(x_F,p_\perp  )$. 
(There is no factor $2$ 
in front of  $N^{nl}$  and $P_{MB}$, since both of them refer to 
quantities  summed over ``left'' and ``right,''  
thus, they are already  twice the unpolarized number densities measured 
on one side exclusively.) 
$N^{nl}(x_F,p_\perp )$ is the number density due to  
nonleading particle production and is obtained by fitting the 
$K^-$ cross section as discussed above. 

The asymmetries for $\pi^+$, $\pi^-$,  
and $\pi^0$ production are then given by  
\begin{eqnarray} 
   A_N^{\pi^+}(x_F) & = &  
    \frac{C[\frac{2}{3}P_{\pi^+ n}(x_F) 
         -\frac{1}{3}P_{\pi^+ \Delta^{0}}(x_F)]}{ 
           N^{nl}(x_F) + P_{\pi^+ n}(x_F)  
           +P_{\pi^+ \Delta^{0}}(x_F)} ;\\  
   A_N^{\pi^-}(x_F) &  = &  \frac{ -\frac{1}{3} C 
   P_{\pi^-\Delta^{++}}(x_F)}{ 
          N^{nl}(x_F ) + 
         P_{\pi^-\Delta^{++}}(x_F)};  \\
   A_N^{\pi^0} (x_F) &  = &  
   \frac{C[\frac{2}{3}  P_{\pi^0 p}(x_F) - \frac{1}{3}  
                P_{\pi^0\Delta^+}(x_F)]}{  
               N^{nl}(x_F)+ P_{\pi^0 p}(x_F) + 
        P_{\pi^0\Delta^+}(x_F)} \,\,.  
\end{eqnarray} 
Here, we dropped the $p_\perp$ dependence of 
$P_{MB}(x_F)$ for simplicity. For the parameter $C$, we use the value $0.6$. 
The results, calculated for $p_\perp =0.75$ GeV/c,  
are shown in Fig. 1d and are compared to 
 the data of the E704 Collaboration \cite{FNAL}.

In the (relativistic) meson-cloud model, the probabilities 
for different angular momenta are not connected by Clebsch-Gordan 
coefficients. They have to be calculated explicitly.  
For a $MB(\frac{1}{2})$ fluctuation, we obtain 
\begin{equation}
 |{\cal T}_{BM}^{Ss}|^2 =  \left[(p.P)-m_Nm_B\right]\left[1 +  s.S\right]
      -(p.S)(P.s)   .
\end{equation}
(Note, that $|{\cal T}^{Ss}_{BM}(y)|^2=|{\cal T}^{Ss}_{MB}(1-y)|^2$.) 
Here, $P$, $S$  and $p$, $s$ are the four-momenta and spin  
of the nucleon and the baryon, respectively.    
We choose the $z$-axis as polarization axis and $x$ as ``longitudinal 
direction'' and consider the  case $p_z=0$ which is relevant for 
the left-right asymmetry.  
Then, in the infinite momentum frame, 
IMF, the four-momenta and transverse spins vectors can be 
parametrized as  $P=(P_L+m_N^2/2P_L,P_L,0,0)$, 
$p=(yP_L+ (m_B^2+p_\perp^2)/2yP_L,yP_L,p_\perp ,0)$, 
$S=(0,0,0,1)$ and $s=(0,0,0,1)$.   
We immediately see that 
${\cal T}_{BM}^{\frac{1}{2}\frac{1}{2}}=0$ and 
${\cal T}_{BM}^{\frac{1}{2}-\frac{1}{2}}$ is given by 
\begin{equation} 
   |{\cal T}_{BM}^{\frac{1}{2}-\frac{1}{2}}|^2 =\frac{1}{y} 
\left[(m_Ny-m_B)^2 +p_\perp^2 \right] . 
\end{equation} 
The orbital angular 
momentum component $L_z$ must be positive ($+1$) in order to 
compensate  the spin of the baryon.   
Thus, the left-right asymmetry is positive  
for $\pi^+$ and also for $\pi^0$.

For spin-$\frac{3}{2}$ baryons, the amplitudes can be obtained by 
calculating the trace over the  nucleon 
and the Rarita-Schwinger spin-vectors, $u_\alpha$, for the 
baryon 
\begin{equation} 
 |{\cal T}_{BM}^{Ss}|^2 = {\rm Tr} \left[ u(P,S) \bar u(P,S)\
                    u_{\alpha}(p,s) \bar u_{\beta}(p,s)
             \right]
(P - p)^{\alpha} (P - p)^{\beta},
\label{trace} 
\end{equation}
$u_{\alpha}(p,s)=\sum_m  C(\frac{3}{2} s 
| 1 m; \frac{1}{2} s-m)
    \epsilon_{\alpha} (m) u(p,s-m)$ are the  Rarita-Schwinger 
spin-vector $u_\alpha$ for the baryon. 
The polarization vectors are 
$\epsilon (m) =(\frac{\bme_m .{\bf p}}{m_B}, 
\bme_m + \frac{{\bf p} (\bme_m.{\bf p})}{ 
m_B(p_0+m_B)})$ and we use the representations 
$\bme_{\pm 1} =\mp (1,\pm i,0)/\sqrt{2}$ 
and $\bme_0 = (0,0,1)$ in the rest frame of the $\Delta$.  
We have $P.\epsilon (0)=0$ and only 
two  amplitudes are nonzero. We obtain 
$|{\cal T}_{BM}^{\frac{1}{2}-\frac{1}{2}}|^2= \frac{1}{3}
   |{\cal T}_{BM}^{\frac{1}{2}\frac{3}{2}}|^2$ and  
\begin{equation}
   |{\cal T}_{BM}^{\frac{1}{2}\frac{3}{2}}|^2 =  \frac{1}{8y^3m_B^2}
    \left[(m_Ny+m_B)^2+p_\perp^2 \right]^2
   \left[(m_Ny-m_B)^2+p_\perp^2 \right]\,\,. 
\label{baryon} 
\end{equation} 
Since the total orbital angular momentum of the meson-baryon system 
is $\mp 1$ for the 
transitions $|{\cal T}_{BM}^{\frac{1}{2}\frac{3}{2}}|^2$  
and $|{\cal T}_{BM}^{\frac{1}{2}-\frac{1}{2}}|^2$, respectively, and  
$|{\cal T}_{BM}^{\frac{1}{2}\frac{3}{2}}|^2 > 
|{\cal T}_{BM}^{\frac{1}{2}-\frac{1}{2}}|^2$, 
the left-right asymmetry is negative for $\pi^-$ production. 

Collecting all factors and defining the unpolarized 
probabilities as $P_{MB}(y)$, the 
left-right asymmetries $A_N$ in the meson-cloud model can be expressed as   
\begin{eqnarray}
  A_N^{\pi^+}(x_F)
 & = &\frac{C [P_{\pi^+ n}(x_F)
-\frac{1}{2} P_{\pi^+\Delta^{0}}(x_F) ]}
 {N^{nl}(x_F) + P_{\pi^+n}(x_F )
  +P_{\pi^+\Delta^{0}}(x_F)}; \nonumber\\
  A_N^{\pi^-}(x_F)
 & = & \frac{-\frac{1}{2}C P_{\pi^-\Delta^{++}}(x_F)}
 {N^{nl}(x_F) + P_{\pi^-\Delta^{++}}(x_F)};
  \nonumber \\ 
  A_N^{\pi^0}(x_F) 
 & = &\frac{C [P_{\pi^0 p}(x_F)
-\frac{1}{2} P_{\pi^0\Delta^{+}}(x_F)]}
 {N^{nl}(x_F) + P_{\pi^0p}(x_F)
  +P_{\pi^0\Delta^{+}}(x_F)}.  
\end{eqnarray} 
Note, that the probabilities for the special case, $p_z=0$, 
are related to  each other by ``Clebsch-Gordan-like'' expressions.  
The result we obtain for the asymmetries is shown in Fig. 2.d.  
Since  $P_{\pi\Delta}$ 
peak at larger $y$ values than $P_{\pi N}$, the asymmetries for 
$\pi^+$ and $\pi^0$ turn down at high $x_F$. 
To confirm this relativistic effect  data at higher $x_F$ is 
needed. 


In conclusion, we have 
shown that energetically favored light-cone meson-baryon 
fluctuations cannot only account for the leading meson 
spectrum in inclusive meson production at moderate 
transverse momenta, but also 
explain the substantial  left-right asymmetries 
measured in such processes using 
transversely polarized proton projectiles and unpolarized proton targets. 
Left-right asymmetries, thus, reflect the correlation between total and 
orbital angular momenta of energetically favored 
light-cone meson-baryon fluctuations.

The author would like to thank Liang Zuo-tang and Meng Ta-chung for 
helpful discussions at an early stage of this work.   
This work is supported  by the Australian Research Council.

\begin{figure}
\epsfig{figure=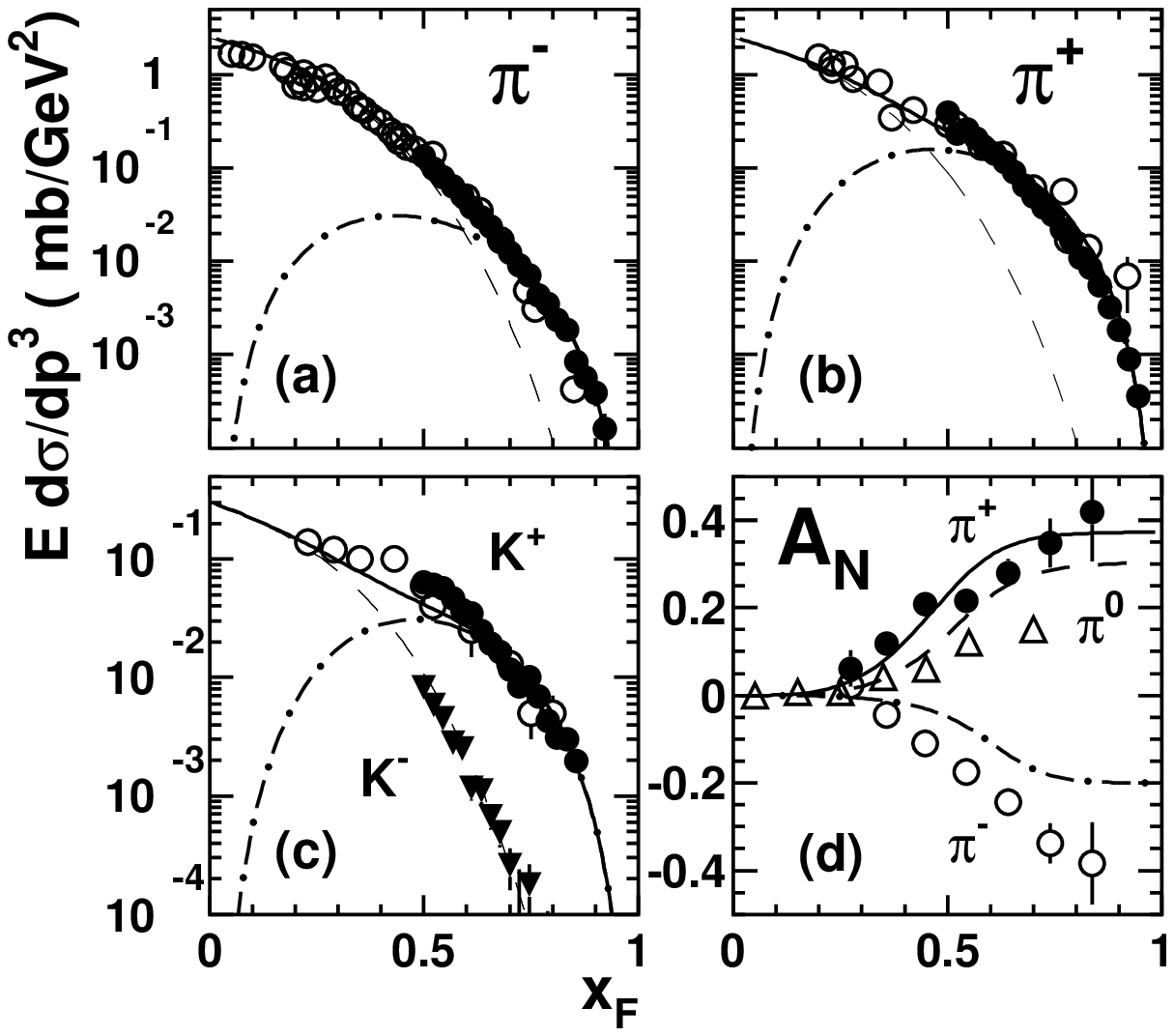,height=12.cm}
\vspace*{-1.cm} 
\caption{The invariant cross section $Ed\sigma/d^3p$ 
for $\pi^-$ (a), $\pi^+$ (b) and $K^\pm$ (c) production as a function 
of $x_F$. The dash-dotted curves represent the calculated contributions  of 
leading particle production.   
The dashed curves stand for the spectra of the nonleading 
particles obtained by fitting the $K^-$ distribution. 
The full curves are the sum of the leading and nonleading 
contributions. 
The solid and open circles are for the $p_\perp =0.75$ GeV/c and 
$p_\perp =0.8$ GeV/c data points, respectively. 
The $0.8$ GeV/c data have been normalized to the 
$0.75$ GeV/c data to account for the difference between the 
transverse momenta in the two experiments. 
The data are taken from Ref.[20]. 
(d) The calculated left-right asymmetries; 
the solid, dashed and dash-dotted curves are for 
$\pi^+$, $\pi^0$ and $\pi^-$, respectively.   
The data are taken from Ref.[1].} 
\end{figure}  

\begin{figure}
\epsfig{figure=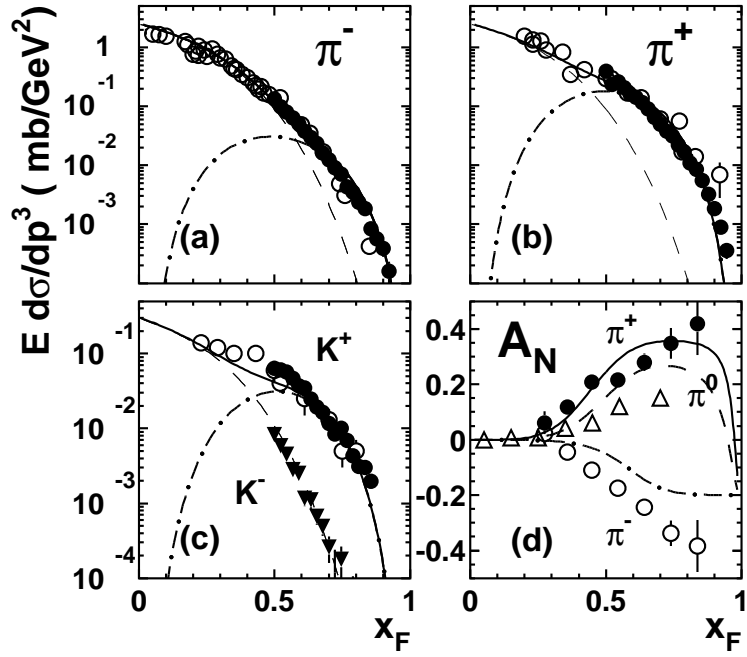,height=12.cm}
\vspace*{-1.cm} 
\caption{The same as in Fig.1 calculated in the meson-cloud model. \
The parameter $C$ is set to $0.4$. }
\end{figure}  

\end{document}